\newcommand{\BF}[1]{\mbox{\boldmath $#1$}}

\documentstyle[aps,prl,multicol]{revtex}

\def\ins#1{}

\def\iems#1{}
\def\comment#1{}

\def\cm#1{}

\begin{document}
\setcounter{figure}{0}
\Roman{figure}

\title{{
Towards a Simulation of Quantum Computers by Classical Systems
}}
\author{Z. Haba%
 \thanks{On leave from Institute of Theoretical Physics, University of
Wroclaw, Poland; e-mail: zhab@ift.uni.wroc.pl}
 and H. Kleinert
 \thanks{Email: kleinert@physik.fu-berlin.de \hfil \newline
URL:
http://www.physik.fu-berlin.de/\~{}kleinert \hfill
} }
\address{Institut f\"ur Theoretische Physik,\\
Freie Universit\"at Berlin, Arnimallee 14,
14195 Berlin, Germany}

\maketitle
\begin{abstract}
We present
a two-dimensional
classical stochastic differential equation
for a displacement field
of a point particle
in
two dimensions
and show that  its components define
real and imaginary parts of a complex field
satisfying
the
Schr\"odinger equation
of a harmonic oscillator.
In this way we
derive
the discrete
oscillator spectrum
from classical dynamics.
The model is then generalized
to an arbitrary potential.
This opens up the possibility of
efficiently simulating quantum
computers with the help of
classical systems.

\end{abstract}
\begin{multicols}{2}

%
%

~\\
{\bf 1.} In a recent analysis
of quantum mechanics from
the point of view of information processing,
one of us
\cite{1Hooft}
pointed out that decoherence will become an insurmountable
obstacle for the practical construction of quantum computers.
It was suggested that instead of
relying on quantum behavior of microparticles
it seems more promising
to simulate quantum behavior with the help of
fast classical systems.
As a step towards such a goal
we construct
a
simple classical model
which allows us to simulate
the quantum behavior of a harmonic oscillator.
In particular we show that
the discrete energy spectrum
with a definite ground state energy can be obtained in a classical
model.
In the latter respect we go
beyond an earlier model
in Ref.~\cite{2Hooft}
whose spectrum had the defect of being unbounded from below.
Finally, the model is generalized to an arbitrary
potential.

~\\
{\bf 2.} For a point particle
in two dimensions
we define a time-independent
displacement field
${\bf u}({\bf x}) =\left( u^1({\bf x}), u^2({\bf x})\right)$
parametrized by the spatial coordinates ${\bf x}=(x^1,x^2)$.
The
reparametrization freedom
is fixed by
choosing harmonic coordinates
in which
\begin{eqnarray}
  \Delta {\bf u}({\bf x}) = 0,
\label{1}\end{eqnarray}
where $ \Delta $ is the Laplace operator.
This condition
implies that the components $u^1({\bf x})
$
and
$u^2({\bf x})$
satisfy
the Cauchy-Riemann equations
\begin{eqnarray}
 \partial_\mu u^ \nu  = \epsilon_{\mu}{} ^\rho
  \epsilon^{ \nu}{}_{\sigma}
  \partial_ \rho u^ \sigma  ,~~~(\mu, \nu ,\dots=1,2),
\label{2}\end{eqnarray}
 where $ \epsilon _{\mu \nu }$ is the antisymmetric
Levi-Civita pseudotensor. The metric is $ \delta _{\mu \nu }$,
so that indices can be sub- or superscripts.

The
particle is supposed to be in contact with a
heat bath of ``temperature"
$ \hbar  $. Its classical orbits ${\bf x}(t)$
are  assumed to follow
 a stochastic
 differential equation
consisting of a fixed rotation and a
random translation in
the diagonal direction ${\bf n}\equiv (1,1)$:
\begin{eqnarray}
  \dot {\bf x}(t) & = & \BF\omega \times {\bf x}(t)
 +{\bf n}\,{ \eta }(t),
\label{4}\end{eqnarray}
where
${\BF  \omega }$
is the rotation vector of length $ \omega $
pointing orthogonal to the plane,
and ${ \eta }(t)$
 a
white-noise  variable
with zero expectation and
the correlation function
\begin{eqnarray}
   \langle  \eta (t)  \eta  (t') \rangle
 =\hbar \,\delta (t-t').
\label{10eta}\end{eqnarray}
For a particle orbit ${\bf x}(t)$
starting at ${\bf x}(0)={\bf x}$, the position
 ${\bf x}(t)$ at a later time $t$
is a function of ${\bf x}$ and a
{\em functional\/} of the
the noise variable ${\eta }(t')$ for $0<t'<t$:
\begin{equation}
 {\bf x}(t)={\bf X}_t[{\bf x};{\eta}].
\label{@Xt}
\end{equation}
To simplify the notation we indicate the time dependence
of functionals of $ \eta $ by a subscript $t$.

We now introduce a time-dependent
displacement field
$ {\bf u} ({\bf x};t)$
which at $t=0$ is equal to ${\bf u}({\bf x})$ and
evolves with time
as follows:
\begin{eqnarray}
 {\bf u} ({\bf x};t)=
 {\bf u}_t [{\bf x};{\eta}] \equiv
{\bf u} \left({\bf X}_t[{\bf x};{\eta}]\right),
\label{3}\end{eqnarray}
 where the notation
${\bf u}_t [{\bf x};{\eta}] $ indicates the
variables as in
(\ref{@Xt}).

As a consequence of the dynamic equation
(\ref{4}),
the
change of the displacement field in a small time interval
from $t=0$ to $t=\Delta t$
has the expansion
\begin{eqnarray}
 \!\!\Delta  {\bf u}_0[{\bf x};{\eta}]
&=&    \Delta t
\left\{
\left[
\BF \omega \times {\bf x}
\right]
 \cdot\BF\nabla
\right\}
{\bf u} _0[{\bf x};{\eta}]
\nonumber \\[1mm]
&&
\!\!\!\!\!\!\!\!\!\!\!\!\!\!\!\!\!\!\!\!\!\!\!\!\!\!\!\!\!
+~\int _0^{ \Delta t}\!\!\! dt' \,
    {\eta }(t')\,\left({\bf n} \cdot\BF\nabla\right)
{\bf u} _0[{\bf x};{\eta}]
 \label{5}\\&&\!\!\!\!\!\!\!\!\!\!\!\!\!\!\!\!\!\!\!\!\!\!\!\!\!\!\!\!\!
+~\frac{1}{2}
\int _0^{\Delta t}\!\!\! dt' \!
\int _0^{ \Delta t}\!\!\! dt''\,
 \eta (t') \eta (t'')
\left({\bf n}
 \cdot\BF\nabla\right)^2
{\bf u} _0[{\bf x};{\eta}]  +\dots~
.\nonumber
\end{eqnarray}
The omitted terms are of order $  \Delta t^{3/2}$.

We now perform the noise average
of
Eq.~(\ref{5}), defining
the average displacement field
\begin{equation}
\bar{\bf u}({\bf x};t)\equiv \langle {\bf u}_t[{\bf x};{\eta}]\rangle.
\label{@}\end{equation}
Using the vanishing average of $ \eta (t)$ and the correlation
function (\ref{10eta}), we obtain
in the limit $ \Delta t\rightarrow 0$ the
time derivative
\begin{eqnarray}
 \partial_t \bar {\bf u}({\bf x};0)
=
\hat {\cal H}\,
\bar{\bf u} ({\bf x};0)
\label{12}\end{eqnarray}
with the
time displacement operator
\begin{eqnarray}
 \hat{\cal H}
\equiv
\left\{ \left[
{\BF \omega }\times {\bf x}\right] \cdot{\BF\nabla}\right\}
+
\frac{\hbar }{2}
({\bf n}\cdot {\BF  \nabla})^2
.
\label{12o}\end{eqnarray}
The average
displacement field
$
\bar{\bf u}({\bf x};t)$ at an arbitrary time $t$ is obtained by
the operation
\begin{equation}
\bar{\bf u}({\bf x};t)=
\hat {\cal U}(t)
\bar{\bf u}({\bf x};0)
\equiv e^{\hat{\cal H}
 t}
\bar{\bf u}({\bf x};0).
\label{@}\end{equation}
 Note that the average over $ \eta $
makes the operator $\hat{\cal H}
 $
time-independent: $
\hat{\cal H} \,\hat {\cal U}(t)=
\hat {\cal U}(t)
\hat{\cal H}.$
Moreover, the operator
 $\hat{\cal H}$ commutes with the
Laplace operator $ \Delta $, thus ensuring that the harmonic
property (\ref{1}) of ${\bf u}({\bf x})$ remains true for all times, i.e.,
\begin{eqnarray}
  \Delta {\bf u}({\bf x};t) \equiv 0,
\label{1bg}\end{eqnarray}

~\\
{\bf 3.}
We now show that
Eq.~(\ref{12}) describes the quantum mechanics
of a harmonic oscillator
Let us restrict our attention
to the line
with arbitrary $x^1\equiv x$
and
 $x_2=0$.
Applying the Cauchy-Riemann equations
(\ref{2}),  we can  rewrite Eq.~(\ref{12})
in the pure $x$-form
\begin{eqnarray}
 \partial _t \bar{ u}_t^1  & = & ~~ \omega\,
 x\,\partial _x \bar{ u}_t^2 -
 \frac{ \hbar  }2
 \partial _x^2 \, \bar u_t^2,\label{14a}\\
 \partial _t \bar{u}_t^2  & = & - \omega\, x\,\partial _x \bar{ u}^1_t +
 \frac{ \hbar  }2
 \partial _x^2  \,\bar u^1_t .
\label{14}\end{eqnarray}
 Now we introduce a complex field
\begin{eqnarray}
 \psi(x;t)
 \equiv e^{-  \omega {x^2}/{2\hbar} }\left[
 \bar u^1 \left({ x},t\right)+i
 \bar u^2 \left({ x},t\right)
 \right] ,
\label{16}\end{eqnarray}
where we have written $
 \bar u^\mu_t \left({x}\right)$
 for
 $\bar u^\mu_t \left({\bf x}\right)
| _{x_1=x,x_2=0}$.
 This
satisfies the differential equation
\begin{eqnarray}
 i \hbar \partial _t \psi(x;t) =
\left(- \frac{\hbar ^2}{2} \partial _x^2+\frac{ \omega ^2}{2}x^2-\frac{\hbar  \omega }{2}\right)  \, \psi(x;t),
\label{14S}\end{eqnarray}
which
 is
 the Schr\"odinger
equation
of a harmonic oscillator
with the
discrete  energy spectrum $E_n=(n+1/2)\hbar  \omega $, $n=0,1,2,\dots~$.
%

~\\
{\bf 5.} The
method can easily be generalized
to an arbitrary potential.
We simply replace
(\ref{4})
by
\begin{eqnarray}
  \dot { x}{^1}(t) & = & -\partial _2S^1({\bf x}(t))
 +{n^1}\,{ \eta }(t),\nonumber \\
  \dot { x}{^2}(t) & = & -\partial _1S^1({\bf x}(t))
 +{n^2}\,{ \eta }(t),
\label{4g}\end{eqnarray}
where
${\bf S}({\bf x})$ shares with
${\bf u}({\bf x})$ the harmonic property (\ref{1}):
\begin{equation}
 \Delta {\bf S}({\bf x})=0,
\label{@}\end{equation}
i.e.,
the
functions $S^\mu({\bf x})$
with $\mu=1,2$ fulfill
Cauchy-Riemann equations like
$u^\mu({\bf x})$
in
(\ref{2}).
Repeating the above steps
we find, instead of the operator  (\ref{12o}),
\begin{eqnarray}
 \hat{\cal H}
\equiv
-(\partial _2S^1)\partial _1-
(\partial _1S^1)\partial _2
+
\frac{\hbar }{2}
({\bf n}\cdot {\BF  \nabla})^2
,
\label{12og}\end{eqnarray}
and Eqs.~(\ref{14a}) and (\ref{14}) become:
\begin{eqnarray}
 \partial _t \bar{ u}_t^1  & = & ~~(\partial _xS^1)
\partial _x \bar{ u}_t^2 -
 \frac{ \hbar  }2
 \partial _x^2 \, \bar u_t^2,\label{14ag}\\
 \partial _t \bar{u}_t^2  & = & - (\partial _xS^1)\,\partial _x \bar{ u}^1_t +
 \frac{ \hbar  }2
 \partial _x^2  \,\bar u^1_t .
\label{14g}\end{eqnarray}
This time evolution preserves the harmonic nature of ${\bf u}({\bf x})$.
Indeed, using the harmonic property
$ \Delta {\bf S}({\bf x})=0$
we can easily derive
the following time dependence of the
Cauchy-Riemann combinations in Eq.~(\ref{2}):
\begin{eqnarray}
&&\partial _t(\partial _1u_1-\partial _2u_2)=\hat {\cal H}
(\partial _1u_1-\partial _2u_2)
\nonumber \\
&&~~~-\partial _2\partial _1S^1(\partial _1u_1-\partial _2u_2)
+\partial _2^2S^1(\partial _2u_1+\partial _1u_2),
\nonumber \\
&&\partial _t(\partial _2u_1+\partial _1u_2)=\hat {\cal H}
(\partial _2u_1+\partial _1u_2)\nonumber \\
&&~~~-\partial _2\partial _1S^1(\partial _2u_1+\partial _1u_2)
-\partial _2^2S^1(\partial _1u_1-\partial _2u_2).
\label{@}\end{eqnarray}
Thus $\partial _1u_1-\partial _2u_2$
and
$\partial _2u_1+\partial _1u_2$ which are zero at
any time remain zero for all times.

On account of Eqs.~(\ref{14g}),
 the combination
\begin{eqnarray}
 \psi(x;t)
 \equiv e^{-  S^1({x})/{\hbar} }\left[
 \bar u^1 \left({ x};t\right)+i
 \bar u^2 \left({ x};t\right)
 \right].
\label{16g}\end{eqnarray}
satisfies the Schr\"odinger equation
\begin{eqnarray}
 i \hbar \partial _t \psi(x;t) =
\left[- \frac{\hbar ^2}{2} \partial _x^2+V(x)\right]  \, \psi(x;t),
\label{14Sg}\end{eqnarray}
where the
potential
is related to $S^1(x)$ by the Riccati
differential equation
\begin{equation}
V(x)=
\frac{1}{2}
[\partial _x\,S^1( x)]^2
-\frac{\hbar }{2}
\partial ^2_x\,S^1({x}).
\label{@}\end{equation}
\comment
{
The operator
$\hat{\cal H}$ in
(\ref{12og})
commutes with the Laplacian
in front of any harmonic function,
 thus ensuring the harmonic
property (\ref{1bg}) for all times.
The commutation rule is
\begin{eqnarray}
[{\cal H},\Delta ]&=&
2(\partial _1\partial_2S^1)\Delta
+2(\Delta S^1)\partial _1\partial _2
\\&+&\partial _2(\Delta S^1)]\partial _1
+[\partial _1(\Delta S^1)]\partial _2.
\label{@}\end{eqnarray}
}

The harmonic oscillator
is recovered for the pair of functions
$S^1({\bf x})+
iS^2({\bf x})=  \omega (x^1+ix^2)^2/2$.

~\\
{\bf 5.}
The noise $ \eta (t)$ in the stochastic differential equation
 Eq.~(\ref{4g})
can also be replaced
by a source composed of
deterministic
classical oscillators $q_k(t)$, $k=1,2,\dots\,$
with the equations of motion
\begin{eqnarray}
\dot q_{k}=p_{k},~~~~~
\dot p_{k} =-\omega_{k}^{2}q_{k}
\label{@}\end{eqnarray}
as
\begin{equation}
 \eta (t)\equiv   \sum_{k} \dot q_{k}(t),
\label{@}\end{equation}
The
 initial positions $q_k(0)$ and momenta $p_k(0)$
 are assumed to be
randomly  distributed
with a Boltzmann factor
$e^{- \beta  H_{\rm osc}/ \hbar } $, such that
\begin{equation}
\langle q_k(0)q_k(0)\rangle=\hbar / \omega _k^2,  ~~~~
\langle p_k(0)p_k(0)\rangle=\hbar .
\label{@}\end{equation}
Using the equation of motion
\begin{equation}
\dot  q_k(t)= \omega _k q_k(0) \sin  \omega _k t+p_k(0)\sin  \omega _k t,
\label{@}\end{equation}
we find
the correlation function
\begin{eqnarray}
\langle \dot  q_k(t)\dot q_k(t')\rangle &=& \omega _k^2
\cos  \omega _kt
\cos  \omega _kt'
\langle q_k(0)q_k(0)\rangle
\nonumber \\&&~+
\sin  \omega _kt\sin  \omega _kt
\langle p_k(0)p_k(0)\rangle
\nonumber \\&=&\cos  \omega _k(t-t').
\label{@CF}\end{eqnarray}
We may now assume that the oscillators
$q_k(t)$  are the Fourier components a massless
field, for instance the gravitational field whose frequencies are
$ \omega _k=k$, and
whose random intial conditions
are caused by the big bang.
If the
 sum over $k$ is simply a momentum integral, then (\ref{@CF})
yields a white-noise correlation function
(\ref{10eta})
for $ \eta (t)$.

\comment{
Introducing  the combinations
 of coordinates
 $x_+\equiv x_{1}+x_{2}, x_-\equiv x_{1}-x_{2}$
in the Langevin equations (\ref{4g}), we
 replace
them by the deterministic equations
\begin{equation}
\dot x_+=-2\partial_{+}S^{1}+2\eta,~~~~~\dot x_-=2\partial_{-}S^{1}.
\label{@}\end{equation}
The thermal distribution of the initial data
produces now the quantum-mechanical
uncertainty if $ \beta $ is chosen appropriately
with a factor $1/\hbar $.}

~\\
{\bf 6.} We have shown that it is
possible to simulate
the quantum-mechanical
wave functions
$\psi(x,t)$
 and the energy spectrum
of an arbitrary potential problem
by
classical stochastic equations of motion,
or by  deterministic
equations with random initial conditions.

It remains to solve the open problem
of finding a classical
origin
of the second
important ingredient of quantum theory:
the theory of quantum measurement
to be extracted from the wave function
$\psi(x,t)$.
Only then
shall  we understand how
God throws his dice.

~\\
~\\
Acknowledgement:\\
The authors thank Gerard 't Hooft for
many useful discussions.

\end{multicols}


\begin{thebibliography}{11}
 \bibitem{1Hooft}
 G.~t'Hooft,
hep-th/9903088,
hep-th/003005.

%
\bibitem{2Hooft}
G.~t'Hooft,
Found. Phys. Lett. {\bf 10}, 105 (1997) (quant-ph/9612018).
%
\end{thebibliography}
\end{document}